\documentclass[journal abbreviation, manuscript]{copernicus}
\usepackage{subfigure}
\usepackage{booktabs}
\usepackage{bbm}
\usepackage{placeins}
\usepackage{threeparttable}

\usepackage{bm}

\newcommand{\bx}{\bm{\mathrm{x}}}
\newcommand{\bz}{\bm{\mathrm{z}}}
\newcommand{\bu}{\bm{\mathrm{u}}}
\newcommand{\bY}{\bm{y}}

\newcommand{\design}{\bm{\mathrm{X}}}
\newcommand{\designP}{\bm{\mathrm{X}}^{\prime}}
\newcommand{\deepF}{\mathrm{F}}
\newcommand{\deepU}{\bm{\mathrm{U}}}

\newcommand{\fv}{\bm{f}}
\newcommand{\bk}{\bm{K}}
\newcommand{\bK}{\mathrm{\bk}}
\newcommand{\bS}{\bm{\mathrm{S}}}
\newcommand{\induZ}{\bm{Z}}
\newcommand{\mean}{\bm{\mu}}
\newcommand{\betas}{\bm{\beta}}
\newcommand{\phis}{\bm{\phi}}
\newcommand{\xstar}{\bx}
\newcommand{\ystar}{{y}^*}

\begin{document}

\title{Bayesian Emulation of Multi-fidelity Earth System Modelling Using Hierarchical Gaussian Processes}

\Author[1][x.xiong@exeter.ac.uk]{Xiaoyu}{Xiong} 
\Author[2]{Louise}{Kimpton}
\Author[3]{Huiyi}{Yang}
\Author[4]{Mian}{Xu}
\Author[1]{James}{Salter}
\Author[1]{Peter}{Challenor}

\affil[1]{Department of Mathematics and Statistics, University of Exeter, UK}
\affil[2]{University of Exeter Medical School, University of Exeter, UK}
\affil[3]{Natural Resources Institute, University of Greenwich, UK}
\affil[4]{Northwest Institute of Eco-Environment and Resources, Chinese Academy of Sciences, China}




\runningtitle{Bayesian Emulation of Multi-fidelity Earth System Modelling}

\runningauthor{Xiong et al.}

\received{}
\pubdiscuss{} 
\revised{}
\accepted{}
\published{}


\firstpage{1}

\maketitle

\makeatletter
\@ifundefined{nolinenumbers}{}{\nolinenumbers}
\makeatother

\begin{abstract}
Multi-fidelity Earth system models provide simulations at different levels of complexity and computational cost, but exhaustive exploration of the parameter space at the highest fidelity is often prohibitively expensive. Multi-fidelity emulators can reduce this burden by combining abundant lower-fidelity simulations with limited high-fidelity evaluations. We compare four Gaussian-process-based multi-fidelity approaches: the Kennedy--O'Hagan autoregressive model (K\&O), hierarchical kriging (HK), Bayesian hierarchical emulation for multi-level models (BayHEm), and multi-fidelity deep Gaussian processes (MF-DGP). We evaluate the methods using two contrasting applications: a three-fidelity tsunami simulator and a two-fidelity implementation of the Joint UK Land Environment Simulator (JULES). Performance is assessed using leave-one-out predictive accuracy, uncertainty representation, design requirements, and computational characteristics. In the tsunami application, BayHEm gives the lowest normalised root mean square error (NRMSE = 0.031) and highest SCORE (3.025), while MF-DGP performs worse than the single-fidelity baseline when only 10 high-fidelity simulations are available. In the JULES application, MF-DGP gives the lowest NRMSE (0.079) and highest SCORE (3.032), with all 30 held-out high-fidelity observations lying within their nominal 95\% predictive intervals. These contrasting results show that no single multi-fidelity emulator is uniformly superior. Instead, method choice should reflect the complexity of the inter-fidelity relationship, the amount of high-fidelity information available, and the importance placed on predictive accuracy and uncertainty quantification.
\end{abstract}


\section{Introduction}
Complex mathematical models implemented as computer simulators are widely used to study the Earth system, including climate, land-surface, and tsunami processes. Such simulators encode scientific understanding of real-world processes and are especially valuable when physical experimentation is prohibitively expensive or impossible. Advances in process representation and numerical resolution have made these models increasingly sophisticated, but this sophistication often entails substantial computational cost: a single model evaluation can require hours or days. Statistical emulators (also referred to as surrogates) provide fast approximations to expensive simulators and quantify predictive uncertainty at input settings that have not been evaluated directly \citep{kennedy2000predicting,o2006bayesian}. Gaussian-process (GP) emulators are particularly attractive because they provide a flexible non-parametric representation of the simulator response together with an explicit predictive distribution. By replacing repeated simulator evaluations with inexpensive emulator predictions, they enable sensitivity analysis, uncertainty quantification, calibration, and parameter-space exploration that would otherwise be computationally infeasible.

 Often a simulator can be run at different levels of complexity, with versions ranging from the most sophisticated high level code (high-fidelity) to the most coarse level (low-fidelity). These levels may vary by spatial or temporal resolution, or by the inclusion or exclusion of specific modelling components, assumptions, or features. While the highest-fidelity models are typically of primary interest, they are also the most expensive to run. In contrast, cheaper lower-fidelity models - though less detailed - can offer valuable insights, helping to infer some underlying features of the full system. A notable example is atmosphere-ocean general circulation models \citep{flato2014evaluation}, which simulate climate dynamics across the atmosphere, ocean, land, and sea ice. These models can be run at varying resolutions to balance computational cost with the need for detail, allowing for a flexible exploration of climate processes across scales. 

Recent applications demonstrate the growing importance and effectiveness of multi-level emulation approaches in Earth system modelling. \citet{tran2016building} developed a pioneering framework for building traceable climate model hierarchies using multi-level emulators, demonstrating how fast, low-complexity models can be systematically combined with more detailed atmospheric general circulation models to efficiently explore uncertainty in paleoclimate applications. \citet{harvey2018multi} applied Bayesian linear emulation to the NAME (Numerical Atmospheric-dispersion Modelling Environment) volcanic ash transport and dispersion model, linking fast and slow simulator configurations to quantify sensitivity to uncertain parameters during the 2010 Eyjafjallajökull eruption, revealing that combining many evaluations of a computationally faster configuration with relatively few evaluations of a more accurate configuration enables effective parameter sensitivity analysis and uncertainty quantification. More recently, \citet{fletcher2022toward} developed a convolutional neural network-based emulator for the Community Earth System Model (CESM), combining information from two lower-resolution configurations with a limited number of higher-resolution simulations to achieve 20-40\% computational efficiency gains during model calibration. \citet{servera2023multifidelity} further advanced the field by applying multi-fidelity Gaussian process methods to atmospheric radiative transfer models, showing that multi-fidelity approaches can reduce prediction errors by approximately 50\% compared to single-fidelity emulators while maintaining computational efficiency. These studies exemplify how multi-fidelity emulation frameworks can extract maximum value from limited computational budgets while maintaining predictive accuracy across diverse Earth system applications, motivating the need for systematic investigation of different multi-level Gaussian process methodologies.

The aim of this work is to compare Bayesian approaches for emulating multi-fidelity Earth system models, with particular emphasis on prediction at the highest fidelity by combining simulations from several fidelity levels. We consider four GP-based approaches: the Kennedy--O'Hagan autoregressive model, hierarchical kriging, BayHEm, and MF-DGP. The methods differ in how they transfer information and uncertainty between fidelity levels, in their design requirements, and in their capacity to represent nonlinear inter-fidelity relationships. We evaluate them on two contrasting applications: a three-fidelity tsunami propagation simulator and a two-fidelity implementation of JULES. Together, these case studies allow us to examine predictive accuracy, uncertainty representation, computational characteristics, and the practical consequences of model complexity and high-fidelity sample size.

\section{Overview of hierarchical Gaussian process models}
\subsection{The Gaussian process emulator}\label{sec:gpEmu}

Emulation using Gaussian processes (GPs) is a popular modelling choice \citep{williams2006gaussian} largely due to the predictive accuracy and uncertainty quantification (UQ) offered by GPs through their flexible nonparametric formulation. 

Let $f : \mathbb{R}^d \rightarrow \mathbb{R}$ represent a generic computer model. Define $\design = [\bx_{1}, ..., \bx_{N}]$ as a set of $N$ input vectors with size $N \times d$, and let $\fv = f(\design)$ denote $N$ function evaluations at input parameters $\bx_{1}, ..., \bx_{N}$. The GP formulation of $\fv$ implies function values $\fv$ jointly follow a multivariate Normal taking the form
\begin{equation}\label{eq:GP}
\bm{f} \mid \betas, \phis  \sim \mathcal{N}(\mean(\design; \betas), \mathrm{\bk}(\design, \designP; \phis)),
\end{equation} 
where $\mean(\design; \betas)$ denotes the mean vector dependent on hyperparameters $\betas$,  $\bK(\design, \designP; \phis)$ denotes the covariance matrix between $\fv$ and the entries of $\bK(\design, \designP; \phis)$ are specified by a kernel function  $k(\bx, \bx^\prime; \phis)$ depending on hyperparameters $\phis$. In the GP regression setting,  the observations $\bm{y}= [y(\bx_1), \ldots, y(\bx_N)]^{\mathrm{T}}$  are assumed to be Gaussian distributed with mean $\fv$ and covariance $\lambda \bm{\mathrm{I}}$ where $\lambda$ is the variance of the mean zero Gaussian observation error. In this setting, the likelihood $p(\bm{y} \mid \fv)$ and $p(\fv \mid \betas, \phis )$ form a conjugate pair, so $\fv$ can be integrated out of the model leading to 
\begin{equation}
    \bY \mid \betas, \phis, \lambda \sim \mathcal{N}(\mean(\design; \betas), \mathrm{\Sigma}(\design, \designP; \phis,\lambda)),
\end{equation}
 with covariance matrix $\mathrm{\Sigma}(\design, \designP; \phis,\lambda) =  \mathrm{\bk}(\design, \designP; \phis) + \lambda \bm{\mathrm{I}} $. 
Inference on $(\betas, \phis, \lambda)$ can be performed using derivative-based maximum likelihood estimation (MLE) or full posterior inference.  Let $\ystar$ denote predictions at untried locations $\xstar$ , $\mathrm{\Sigma}(\xstar, \xstar^\prime)$ denote the covariance between untried locations, $\mathrm{\Sigma}(\xstar,\design)$  be the cross-covariance between untried locations and input vectors in the training data, and the entries of $\mathrm{\Sigma}(\cdot)$ are evaluated with the covariance function $
  c(\bx, \bx^\prime) = k(\bx, \bx^{\prime}; \phis) +  \lambda \mathbbm{1}(\bx = \bx^\prime)$. 

The property of GPs implies $\ystar$ follows a Gaussian distribution:
\begin{align}\label{eq:gpPredic}
    &\ystar | \design, \bY \sim \mathcal{N}(\mu^*(\bx) , \Sigma^*(\bx, \bx^\prime)) \qquad \quad  \text{with} & \\
   & \mu^*(\xstar) = \mu(\bx) + \Sigma(\bx,\design)\Sigma(\design, \designP)^{-1}(\bY-\bm{\mu}(\design)) \nonumber &\\
   & \Sigma^*(\bx, \bx^\prime) = c(\bx, \bx^\prime) - \Sigma(\bx, \design)\Sigma(\design, \designP)^{-1}\Sigma(\design, \bx) \nonumber &
\end{align} 

For notational simplicity, we omit the hyperparameters $(\betas, \phis, \lambda)$ in equation (\ref{eq:gpPredic}). In the following Sections \ref{sec:k&Omodel}, \ref{sec:hkmodel}, \ref{sec:deepishModel} and \ref{sec:mfDGP}, all hierarchical GP equations maintain the same mean and covariance functions defined here for the fidelity-wise GP. Similarly, we omit all hyperparameters from these equations for brevity. 

In the subsequent sections, we review four hierarchical GP models: (i) the Kennedy and O’Hagan autoregressive model  in Section \ref{sec:k&Omodel}, (ii) the hierarchical kriging (HK) model in Section \ref{sec:hkmodel}, (iii) the deepish GP model in Section \ref{sec:deepishModel}, and (iv) the multi-fidelity deep GP model (MF-DGP) in Section \ref{sec:mfDGP}. Throughout these sections, we consider a  $T$-fidelity computer model with dataset $\mathcal{D}$ containing observations across $T$ fidelity levels. At each fidelity level $t$, we denote input locations by $\design_t = [\bx^t_1,\ldots, \bx^t_{n_t}]$ and the corresponding output observations by $\bY_t (\design_t) = [y^t_1(\bx^t_1), \ldots, y^t_{n_t}(\bx^t_{n_t})]$ where $n_t$ represents the number of input--output pairs at fidelity level $t$. We use the shorthand $\bY_t$ to denote $\bY_t (\design_t)$ for brevity.

\subsection{The Kennedy and O’Hagan autoregressive model } \label{sec:k&Omodel}
The Kennedy and O'Hagan autoregressive model \citep{kennedy2000predicting} represents each fidelity level of a computer code as a Gaussian process. For any two adjacent fidelity levels, the relationship is modelled as:
 \begin{align}\label{eq:k&oMethod}
&\bY_{t} =\rho_{t-1} \bY_{t-1} + \bm\delta_t   \qquad \quad  \text{with} &\\
&\bY_{t-1}  \sim \mathcal{N}(\mean_{t-1}(\design_{t}), \mathrm{\Sigma}_{t-1}(\design_{t}, \designP_{t})) \nonumber &\\
&\bm\delta_t   \sim \mathcal{N}(\mean_{\delta}(\design_{t}), \mathrm{\Sigma}_{\delta}(\design_{t}, \designP_{t})) \nonumber&
\end{align}
Here, $\bm\delta_t = \bm\delta_t(\design_t)$ captures the difference between adjacent fidelity levels. The functions $\bm\mu_{t-1}(.)$ and $\Sigma_{t-1}(.)$ denote the mean and covariance functions at level $t-1$, obtained by training a GP on the lower fidelity data. The parameter $\rho_{t-1}$ is a scaling factor that captures the correlation between the two fidelity levels. The functions $\mean_\delta(.)$ and $\Sigma_\delta(.)$ represent the mean and covariance of the difference term, trained on the differences computed from running the two fidelity levels at shared design points. This formulation allows for the full posterior predictive distribution to be captured, while maintaining the hierarchical autoregressive structure between fidelity levels. For a pair of adjacent fidelity levels, define $\bY = [\bY_{t-1},\bY_t]$ and $\design= [\design_{t-1}, \design_t]$. Under the K\&O model:
\begin{align}
   &\bY  \sim \mathcal{N}(\mean(\design), \mathrm{\Sigma}(\design, \designP)) \qquad \quad  \text{with} &\\
   &\mean(\design) = [\mean_{t-1}(\design_{t-1}), \quad\rho_{t-1}\mean_{t-1}(\design_{t}) + \mean_{\delta}(\design_{t})] \nonumber &\\
   &\mathrm{\Sigma}(\design, \designP) = \begin{bmatrix} \label{eq:covK&O} 
\Sigma_{t-1}(\design_{t-1},\designP_{t-1}) & \rho_{t-1}\Sigma_{t-1}(\design_{t-1},\design_t) \\
\rho_{t-1}\Sigma_{t-1}(\design_t,\design_{t-1}) & \rho^2_{t-1}\Sigma_{t-1}(\design_t,\designP_t) + \Sigma_{\delta}(\design_t, \designP_t)
\end{bmatrix}&
\end{align}
For a new point $\bx$ at level $t$, the posterior predictive distribution follows Equation (\ref{eq:gpPredic}), where $\mu(\bx) = \rho_{t-1} \mu_{t-1}(\bx) + \mu_{\delta}(\bx)$, $\Sigma(\bx, \design) = [\rho_{t-1}\Sigma_{t-1}(\bx,\design_{t-1}), \rho^2_{t-1}\Sigma_{t-1}(\bx,\design) + \Sigma_{\delta}(\bx, \design_t)]$, $\mathrm{\Sigma}(\design, \designP)$ is given in Equation (\ref{eq:covK&O}), and $c(\bx, \bx^\prime) = {\rho_{t-1}}^2 c_{t-1}(\bx, \bx^\prime) + c_{\delta}(\bx, \bx^\prime)$.

Although the derivation is illustrated for a pair of adjacent fidelity levels $(t-1, t)$ for clarity, the model naturally generalises to $T$ fidelity levels, as originally proposed by \cite{kennedy2000predicting}. In this generalised formulation, the highest fidelity model at level $T$ is expressed as a function of all lower-fidelity models. Let $\bY_T$ represent observations at the highest level, the recursive autoregressive structure gives:
\begin{align*}
\bY_T &= \rho_{T-1} \cdot \bY_{T-1} + \bm\delta_{T} \\
&= \rho_{T-1} \cdot (\rho_{T-2} \cdot \bY_{T-2} + \bm\delta_{T-1}) + \delta_{T} \\
&= \rho_{T-1}\rho_{T-2} \cdot \bY_{T-2} + \rho_{T-1}\bm\delta_{T-1} + \bm\delta_{T} 
\end{align*}

Continuing this recursive substitution all the way down to the lowest fidelity level, we get:
\begin{align}
\bY_T &= P_1 \cdot \bY_1 + \sum_{t=2}^{T} P_t \cdot \bm\delta_{t}
\end{align}
where $P_1 = \prod_{i=1}^{T-1} \rho_i $, $P_t = \prod_{i=t}^{T-1} \rho_i$ for $t = 2, \ldots, T-1 $ and $P_T = 1$. Here, $\bY_1$ is a GP representing the lowest fidelity level, and $\bm\delta_{t}$ is an independent GP capturing the  difference at level $t$. The mean and covariance functions of $\bY_1$ and $\bm\delta_{t}$ are as defined in Equation (\ref{eq:k&oMethod}).

\subsection{The hierarchical kriging model}\label{sec:hkmodel}

The hierarchical kriging (HK) model, introduced by \cite{han2012hierarchical}, represents a significant advancement in multi-fidelity surrogate modelling. 
In the HK framework, for two adjacent fidelity levels $(t-1, t)$ , the posterior mean from the lower fidelity level serves as the prior mean for the higher fidelity level, scaled by a factor $\rho_t$  that adjusts the contribution from the preceding level. To extend this to $T$ levels, we define a recursive structure across all fidelity levels. We begin with the lowest fidelity model at level 1:
\begin{equation}
\bY_{1} \sim \mathcal{N}(\mean_{1}(\design_{1}), \Sigma_1(\design_1, \designP_1))
\end{equation}

The predictive distribution for any input $\bx$ at level 1 is given by:
\begin{align}
\ystar_{1} \mid \design_{1}, \bY_{1} &\sim \mathcal{N}({\mu^*}_{1}(\bx), {\Sigma^*}_{1}(\bx, \bx^\prime)) \\
{\mu^*}_{1}(\bx) &= \mu_{1}(\bx) + \Sigma_{1}(\bx,\design_{1})\Sigma_{1}(\design_{1}, \designP_{1
})^{-1}(\bY_{1} - \mu_{1}(\design_{1})) \nonumber \\
{\Sigma^*}_{1}(\bx, \bx^\prime) &= c_{1}(\bx, \bx^\prime) - \Sigma_{1}(\bx, \design_{1})\Sigma_{1}(\design_{1}, \designP_{1})^{-1}\Sigma_{1}(\design_{1}, \bx) \nonumber
\end{align}

For level 2:
\begin{equation}
\bY_{2} \sim \mathcal{N}(\rho_2{\mu^*}_{1}(\design_{2}), \Sigma_2(\design_{2}, \designP_{2}))
\end{equation}

With predictive distribution:
\begin{align}
\ystar_{2} \mid \design_{2}, \bY_2 &\sim \mathcal{N}({\mu^*}_{2}(\bx), {\Sigma^*}_{2}(\bx, \bx^\prime)) \\
{\mu^*}_{2}(\bx) &= \rho_2 {\mu^*}_{1}(\bx) + \Sigma_2(\bx,\design_2)\Sigma_2(\design_2, \designP_2)^{-1}(\bY_2 - \rho_2{\mu^*}_{1}(\design_2)) \nonumber \\
{\Sigma^*}_{2}(\bx, \bx^\prime) &= c_2(\bx, \bx^\prime) - \Sigma_2(\bx, \design_2)\Sigma_2(\design_2, \designP_2)^{-1}\Sigma_2(\design_2, \bx) \nonumber
\end{align}

Generalising to any level $t$ (where $2 \leq t \leq T$):
\begin{equation}
\bY_{t} \sim \mathcal{N}(\rho_t{\mu^*}_{t-1}(\design_{t}), \Sigma_t(\design_{t}, \designP_{t}))
\end{equation}

With predictive distribution:
\begin{align}
\ystar_{t} \mid \design_{t}, \bY_t &\sim \mathcal{N}({\mu^*}_{t}(\bx), {\Sigma^*}_{t}(\bx, \bx^\prime)) \\
{\mu^*}_{t}(\bx) &= \rho_t {\mu^*}_{t-1}(\bx) + \Sigma_t(\bx,\design_t)\Sigma_t(\design_t, \designP_t)^{-1}(\bY_t - \rho_t{\mu^*}_{t-1}(\design_t)) \nonumber \\
{\Sigma^*}_{t}(\bx, \bx^\prime) &= c_t(\bx, \bx^\prime) - \Sigma_t(\bx, \design_t)\Sigma_t(\design_t, \designP_t)^{-1}\Sigma_t(\design_t, \bx) \nonumber
\end{align}

To express the prediction at the highest fidelity level $T$ in terms of all lower levels, we recursively substitute:
\begin{align}
{\mu^*}_{T}(\bx) &= \prod_{i=2}^{T} \rho_i \cdot {\mu^*}_{1}(\bx) + \sum_{t=2}^{T} \left( \prod_{i=t+1}^{T} \rho_i \cdot \Sigma_t(\bx,\design_t)\Sigma_t(\design_t, \designP_t)^{-1}(\bY_t - \rho_t{\mu^*}_{t-1}(\design_t)) \right)
\end{align}
where we define $\prod_{i=T+1}^{T} \rho_i = 1$ for notational convenience.

The predictive variance at level $T$ remains:
\begin{align}
{\Sigma^*}_{T}(\bx, \bx^\prime) = c_T(\bx, \bx^\prime) - \Sigma_T(\bx, \design_T)\Sigma_T(\design_T, \designP_T)^{-1}\Sigma_T(\design_T, \bx)
\end{align}

This generalised formulation reveals that in the HK framework, the prediction at the highest fidelity level $T$ is a weighted combination of the scaled prediction from the lowest fidelity model, scaled by the product of all $\rho$ values and correction terms from each intermediate fidelity level, each scaled by the product of $\rho$ values from higher levels.

Compared to the K\&O model, this hierarchical formulation eliminates the need to explicitly model cross-correlations between fidelity levels. As a result, the HK approach is particularly well-suited for applications where the fidelity hierarchy is primarily characterised by similarities in mean response, while also preserving computational scalability in high-dimensional multi-fidelity settings.

\subsection{Bayesian hierarchical emulators for
multi-level models: BayHEm} \label{sec:deepishModel}
The BayHEm \citep{kimpton2025bayhem} is a method of multi-fidelity emulation that models the highest level of the model using information from all lower levels. Unlike the K\&O model, which fits independent GPs to levels (or level differences) and typically requires nested designs, the BayHEm approach fits a single hierarchical GP without nesting constraints. Compared to the above HK model, in the BayHEm model, both the posterior mean and covariance from level $t-1$ become the prior mean and covariance for level $t$. In other words, the posterior distribution at level $t-1$ serves directly as the prior distribution for level $t$.

The most restrictive two-level BayHEm model is given by the following hierarchical model:
\begin{align}
    \bY_{2}(\design_2) | \betas, \phis, \lambda \sim \mathcal{N}({\mean^*}_{\!\!\!\!\!1}(\design_{2}), {\Sigma^*}_{\!\!\!\!\!1}(\design_{2}, \designP_{2})) \\
    \bY_{1}(\design_1) | \betas, \phis, \lambda \sim \mathcal{N}({\mean}_{0}(\design_{1}), {\Sigma}_{0}(\design_{1}, \designP_{1}))
\end{align}
Following this hierarchical structure, the posterior predictive distribution for any untried location $\bx$ at level $2$ takes the form:
\begin{align}\label{hf_prediction_deepishGP}
\ystar_{2} (\bx) | \ystar_{1}(\bx),\betas, \phis, \lambda \sim \mathcal{N}({\mu^*}_{\!\!\!\!\!2}(\bx) , {\Sigma^*}_{\!\!\!\!\!2}(\bx, \bx^\prime)) \qquad \quad  \text{with} \\
{\mu^*}_{\!\!\!\!\!2}(\bx) = {\mu^*}_{\!\!\!\!\!1}(\bx) +\Sigma^*_{1}(\bx,\design_2)\Sigma^*_{1}(\design_2, \designP_2)^{-1}\big(\bY_2 (\design_2) - {\mean^*}_{\!\!\!\!\!1}(\design_{2})\big) \nonumber \\
{\Sigma^*}_{\!\!\!\!\!2}(\bx, \bx^\prime) = c^*_{1}(\bx, \bx^\prime) - \Sigma^*_{1}(\bx, \design_2)\Sigma^*_{1}(\design_2, \designP_2)^{-1}\Sigma^*_{1}(\design_2, \bx). \nonumber
\end{align}
This can be extended for a model with $T$ levels such that 
\begin{align}
    \bY_{t}(\design_t) | \betas, \phis, \lambda \sim \mathcal{N}({\mean^*}_{\!\!\!\!\!t-1}(\design_{t}), {\Sigma^*}_{\!\!\!\!\!t-1}(\design_{t}, \designP_{t})) \\
    \bY_{1}(\design_1) | \betas, \phis, \lambda \sim \mathcal{N}({\mean}_{0}(\design_{1}), {\Sigma}_{0}(\design_{1}, \designP_{1}))
\end{align}
for $t = 2, \ldots, T$ where the posterior predictive distribution for any untried location $\bx$ at level $t$ takes the form:
\begin{align}\label{hf_prediction_bayhem_general}
\ystar_{t} (\bx) | \ystar_{t-1}(\bx),\betas, \phis, \lambda \sim \mathcal{N}({\mu^*}_{\!\!\!\!\!t}(\bx) , {\Sigma^*}_{\!\!\!\!\!t}(\bx, \bx^\prime)) \qquad \quad  \text{with} \\
{\mu^*}_{\!\!\!\!\!t}(\bx) = {\mu^*}_{\!\!\!\!\!t-1}(\bx) +\Sigma^*_{t-1}(\bx,\design_t)\Sigma^*_{t-1}(\design_t, \designP_t)^{-1}\big(\bY_t (\design_t) - {\mean^*}_{\!\!\!\!\!t-1}(\design_{t})\big) \nonumber \\
{\Sigma^*}_{\!\!\!\!\!t}(\bx, \bx^\prime) = c^*_{t-1}(\bx, \bx^\prime) - \Sigma^*_{t-1}(\bx, \design_t)\Sigma^*_{t-1}(\design_t, \designP_t)^{-1}\Sigma^*_{t-1}(\design_t, \bx) \nonumber
\end{align}
Potential benefits of this approach include fewer hyperparameters to estimate and lower computational cost, since it does not require a nested input design—all while maintaining model efficiency and scalability. The fully recursive model can be made more flexible by specifying alternative hyperparameters and/or functions at selected levels. In particular, you can introduce new hyperparameters only at higher levels to allow differences across fidelities while retaining dependence. For example, you might keep the posterior mean from level $t-1$ as the prior mean at level $t$ whilst updating the covariance structure at level $t$:
\begin{align}
    \bY_{t-1}(\design_{t-1}) | \betas, \phis, \lambda &\sim \mathcal{N}({\mean^*}_{\!\!\!\!\!t-2}(\design_{t-1}), {\Sigma^*}_{\!\!\!\!\!t-2}(\design_{t-1}, \designP_{t-1})) \\
    \bY_{t}(\design_t) | \betas, \phis, \lambda, \phis^{\text{new}}, \lambda^{\text{new}} &\sim \mathcal{N}({\mean^*}_{\!\!\!\!\!t-1}(\design_{t}), {\Sigma^{\text{new}}}(\design_{t}, \designP_{t})). 
\end{align}
Or conversely specifying a new mean while maintaining the form of the covariance. It is also possible to use an additive form, for example:
\begin{equation}
    \bY_{t}(\design_t) | \betas, \phis, \lambda \sim \mathcal{N}({\mean^*}_{\!\!\!\!\!t-1}(\design_{t}) + {\mean^{\text{new}}}(\design_{t}), {\Sigma^*}_{\!\!\!\!\!t-1}(\design_{t}, \designP_{t}) + {\Sigma^{\text{new}}}(\design_{t}, \designP_{t}))
\end{equation}
which preserves the relationship between levels but adds flexibility where needed. This mirrors the K\&O additive approach, but the lower levels are embedded as a prior rather than assuming independence.

\subsection{Multi-fidelity deep Gaussian processes} \label{sec:mfDGP}

The K\&O and HK methods rely on simple linear correlations between fidelities, and BayHEm relaxes this assumption through a strong prior. These approaches share a common limitation: they employ additive hierarchical structures where predictions at level $t$ are constructed as corrections to predictions from level $t-1$, constraining the inter-fidelity relationship to be approximately linear. Multi-fidelity Deep Gaussian Processes (MF-DGP) address this by incorporating deep architectures capable of capturing complex, nonlinear inter-fidelity relationships.

Deep Gaussian Processes (DGPs) extend standard GPs by composing multiple GP layers hierarchically: the output of one layer becomes the input to the next, enabling the model to learn flexible, nonlinear mappings without specifying a parametric form. A single-layer GP is limited by its kernel's expressiveness—once a kernel structure is chosen, the class of functions it can represent is fixed. DGPs overcome this by learning hierarchical representations non-parametrically, where each layer refines the mapping learned by the previous one. This composition of layers allows DGPs to capture increasingly complex functions while retaining the GP properties of principled uncertainty quantification and growing complexity with data. To make DGPs computationally tractable, sparse variational inference is employed, replacing the full $\mathcal{O}(n^3)$ GP computation with an $\mathcal{O}(M^2 n)$ approximation using $M \ll n$ inducing points at each layer. Detailed descriptions of sparse GPs and DGPs are provided in Supplementary Materials Sections S1 and S2, respectively.

MF-DGP adapts this deep architecture to the multi-fidelity setting in two ways. First, the fidelity levels are mapped onto the layers of the deep GP, so that information propagates naturally from lower to higher fidelities through the network. Second, a composite kernel is introduced at intermediate layers that explicitly captures both linear and nonlinear components of the inter-fidelity relationship, with the flexibility to emphasise different components in different regions of input space—a capability that additive correction frameworks lack. The technical details of the MF-DGP formulation, including the ELBO, the composite kernel structure, and the predictive mechanism, are given in Supplementary Materials Section S3.

\section{Model validation}
We compare the performance of the five models at the highest fidelity level $T$ using three complementary metrics: normalised root mean square error (NRMSE), SCORE, and coverage probability. These metrics collectively assess prediction accuracy and uncertainty quantification quality at level $T$.

The NRMSE \citep{le2015cokriging} measures the prediction accuracy of the mean response at level $T$, normalised by the range of observed values:
\begin{equation*}
  \text{NRMSE}_T = \frac{\sqrt{\frac{1}{n_T} \sum_{i=1}^{n_T} \left(\mu^*_{-i}(\bx^T_i) - y^T_i\right)^2}}{\max_{i \in \{1,\ldots,n_T\}} y^T_i - \min_{i \in \{1,\ldots,n_T\}} y^T_i}
\end{equation*}
where $\mu^*_{-i}(\bx^T_i)$ is the leave-one-out (LOO) predictive mean at location $\bx^T_i$, obtained by training the model on all data points at level $T$ except the $i$-th observation, and $y^T_i$ is the true observation at $\bx^T_i$. The normalisation by the data range makes NRMSE comparable across different output scales.

The SCORE metric \citep{kimpton2025cross} evaluates both prediction accuracy and uncertainty quantification at level $T$ by combining the standardised prediction error with the predictive variance:
\begin{equation*}
    \text{SCORE}_T = \frac{1}{n_T} \sum_{i=1}^{n_T} \left[ -\left(\frac{\mu^*_{-i}(\bx^T_i) - y^T_i}{\sqrt{{\Sigma^*}_{-i,T}(\bx^T_i)}}\right)^2 - \log\left({\Sigma^*}_{-i,T}(\bx^T_i)\right) \right]
\end{equation*}
where ${\Sigma^*}_{-i,T}(\bx^T_i) = {\Sigma^*}_{-i,T}(\bx^T_i, \bx^T_i)$ is the LOO predictive variance at $\bx^T_i$. The first term penalises large standardised errors, while the second term (logarithm of predictive variance) penalises overly confident (too small) or overly uncertain (too large) predictions. A well-calibrated model balances these two components.

The coverage probability quantifies the reliability of the uncertainty estimates at level $T$ by measuring the proportion of true observations that fall within the 95\% prediction intervals:
\begin{equation*}
    \text{Coverage}_T = \frac{1}{n_T} \sum_{i=1}^{n_T} \mathbb{I}\left(y^T_i \in \left[\mu^*_{-i}(\bx^T_i) - 1.96\sqrt{{\Sigma^*}_{-i,T}(\bx^T_i)}, \, \mu^*_{-i}(\bx^T_i) + 1.96\sqrt{{\Sigma^*}_{-i,T}(\bx^T_i)}\right]\right)
\end{equation*}
where $\mathbb{I}(\cdot)$ is the indicator function that equals 1 if the condition is true and 0 otherwise.

Lower NRMSE values indicate better prediction accuracy of the mean response at level $T$. Higher SCORE values indicate better overall performance, reflecting both accurate predictions and well-calibrated uncertainties. Coverage probability close to the nominal level of 0.95 indicates that the prediction intervals are appropriately calibrated—values significantly below 0.95 suggest underestimation of uncertainty (overconfident predictions), while values significantly above 0.95 suggest overestimation of uncertainty (overly conservative predictions). Together, these metrics provide a comprehensive assessment of emulator performance at the highest fidelity level.
\subsection{Computational cost}

We compare the methods across three components of computational cost: design requirements, training time, and prediction time. A summary is given in Table~\ref{tab:computational_complexity}.

\paragraph{Design requirements}
K\&O requires nested designs between adjacent fidelity levels—shared input locations must exist between levels $t-1$ and $t$ to estimate the autoregressive relationship $\bY_t = \rho_{t-1}\bY_{t-1} + \bm{\delta}_t$—which increases the total number of simulator evaluations, particularly at lower fidelity levels. HK, BayHEm, and MF-DGP all permit fully non-nested designs, allowing practitioners to allocate simulator runs across levels without requiring shared locations.

\paragraph{Training time}
K\&O, HK, and BayHEm all fit GPs sequentially across levels, each requiring an $\mathcal{O}(n_t^3)$ Cholesky decomposition per optimisation iteration at level $t$. Hyperparameter optimisation via MLE typically converges in $K_{MLE} \approx 10$--$100$ iterations, giving a shared training complexity of
\begin{equation}
\mathcal{O}\left(K_{MLE} \cdot T \cdot n^3\right).
\end{equation}
BayHEm benefits from having fewer hyperparameters than K\&O and HK, as its recursive use of posterior distributions reduces the number of quantities requiring optimisation, potentially lowering $K_{MLE}$ in practice.

MF-DGP employs sparse variational inference with $M$ inducing points ($M \ll n$), replacing the $n_t \times n_t$ matrices with $M \times M$ matrices and training all levels jointly via stochastic optimisation of the ELBO. The per-iteration cost is $\mathcal{O}(S \cdot T^2 \cdot n \cdot M^2)$, where $S$ is the number of Monte Carlo samples for gradient estimation. Although this per-iteration cost can be lower than $\mathcal{O}(T \cdot n^3)$ when $S \cdot T \cdot M^2 \ll n^2$, stochastic variational inference requires substantially more iterations to converge ($K_{SVI} \approx 1{,}000$--$10{,}000$), giving a total training complexity of
\begin{equation}
\mathcal{O}\left(K_{SVI} \cdot S \cdot T^2 \cdot n \cdot M^2\right).
\end{equation}
The ratio $K_{SVI}/K_{MLE}$ is the dominant factor determining relative training time: despite the inducing-point speedup, MF-DGP typically requires substantially longer wall-clock time than the traditional methods.

\paragraph{Prediction time}
For K\&O, HK, and BayHEm, prediction at the highest fidelity level requires recursive evaluation through all lower levels, each involving matrix-vector products with the $n_t \times n_t$ covariance matrix, giving a per-point cost of
\begin{equation}
\mathcal{O}(T \cdot n^2).
\end{equation}
MF-DGP predictions depend only on the variational parameters and inducing point locations—not on the original training data—and propagate $S$ Monte Carlo samples through $T$ layers, yielding a per-point cost of
\begin{equation}
\mathcal{O}(S \cdot M^2 \cdot T).
\end{equation}
When $M \ll n$, MF-DGP predictions are substantially cheaper, which can be advantageous for uncertainty quantification workflows requiring predictions at many locations.

\begin{table}[h]
\centering
\caption{Computational complexity comparison of multi-fidelity emulation methods}
\label{tab:computational_complexity}
\small
\begin{tabular}{lcccc}
\hline
\textbf{Method} & \textbf{Nested} & \textbf{Training Time} & \textbf{Prediction} & \textbf{Iterations} \\
 & \textbf{Design?} & \textbf{Complexity} & \textbf{(per point)} & \textbf{to Converge} \\
\hline
K\&O & Required & $\mathcal{O}(K_{MLE} \cdot T \cdot n^3)$ & $\mathcal{O}(T \cdot n^2)$ & $K_{MLE} \sim 10$--$100$ \\
HK & Not required & $\mathcal{O}(K_{MLE} \cdot T \cdot n^3)$ & $\mathcal{O}(T \cdot n^2)$ & $K_{MLE} \sim 10$--$100$ \\
BayHEm & Not required & $\mathcal{O}(K_{MLE} \cdot T \cdot n^3)$ & $\mathcal{O}(T \cdot n^2)$ & $K_{MLE} \sim 10$--$100$ \\
 & & \textit{(fewer hyperparameters)} & & \textit{(potentially fewer)} \\
MF-DGP & Not required & $\mathcal{O}(K_{SVI} \cdot S \cdot T^2 \cdot n \cdot M^2)$ & $\mathcal{O}(S \cdot M^2 \cdot T)$ & $K_{SVI} \sim 1{,}000$--$10{,}000$ \\
\hline
\end{tabular}
\begin{tablenotes}
\small
\item Note: $n$ = typical number of observations per level; $T$ = number of fidelity levels; $S$ = Monte Carlo samples; $M$ = number of inducing points (typically $M \ll n$); $K_{MLE}$ = iterations for MLE convergence; $K_{SVI}$ = iterations for stochastic variational inference convergence.
\end{tablenotes}
\end{table}

\section{Applications to real data}
\subsection{Application to the tsunami model}
\subsubsection{The tsunami model}
Tsunami propagation is typically modelled using variants of the shallow water equations \citep{seelinger2021high}. The model takes two-dimensional inputs [$x$, $y$], representing displacements (in km) from the true tsunami source in the $x$ and $y$ coordinates, and outputs four quantities of interest: the maximum wave height and the corresponding time at which it is reached for each of the two buoys. The model has three levels of fidelity with increasing accuracy. In the first level, bathymetry is approximated by a depth average over the entire domain; the second level introduces wetting and drying capabilities using smoothed bathymetry data; the third level incorporates the full bathymetry data. The computational costs on a MacBook Pro with 8GB of RAM for a single parameter set are $16.64$s (level 1), $29.88$s (level 2), and $9099.18$s (level 3), respectively. 
\subsubsection{Data}
\begin{figure}[!thb]
\begin{center}
\caption{Three-level tsunami model outputs.  \label{fig:tsunamiData}}
\subfigure[Level 1 training data]{
\resizebox{0.3\textwidth}{!}{
\includegraphics[scale=0.8]{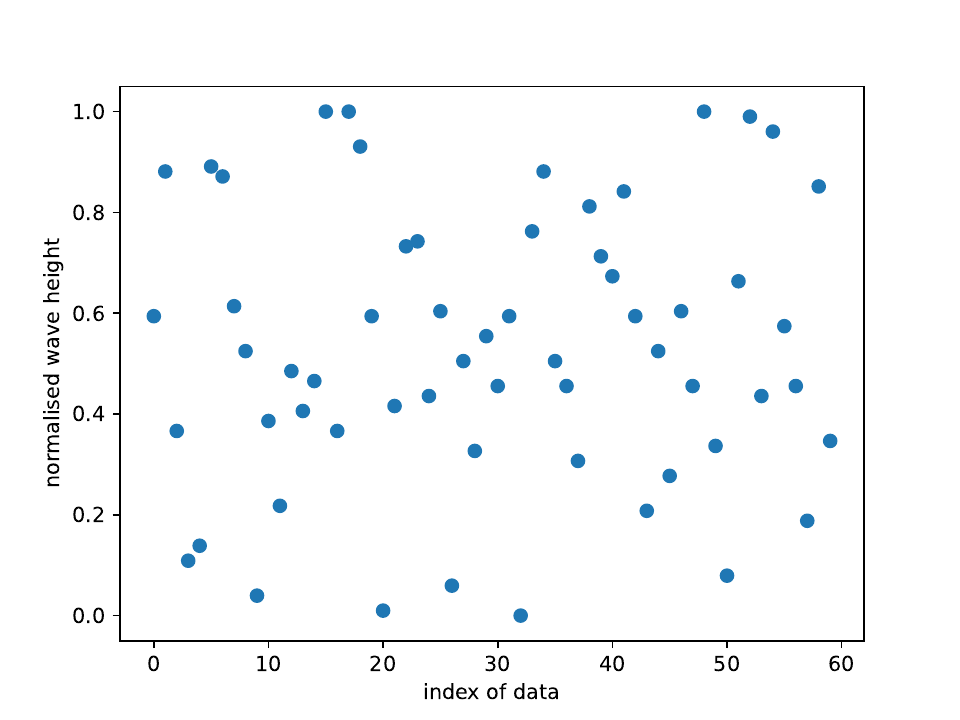}}} 
\subfigure[Level 2 training data]{
\resizebox{0.3\textwidth}{!}{
\includegraphics[scale=0.8]{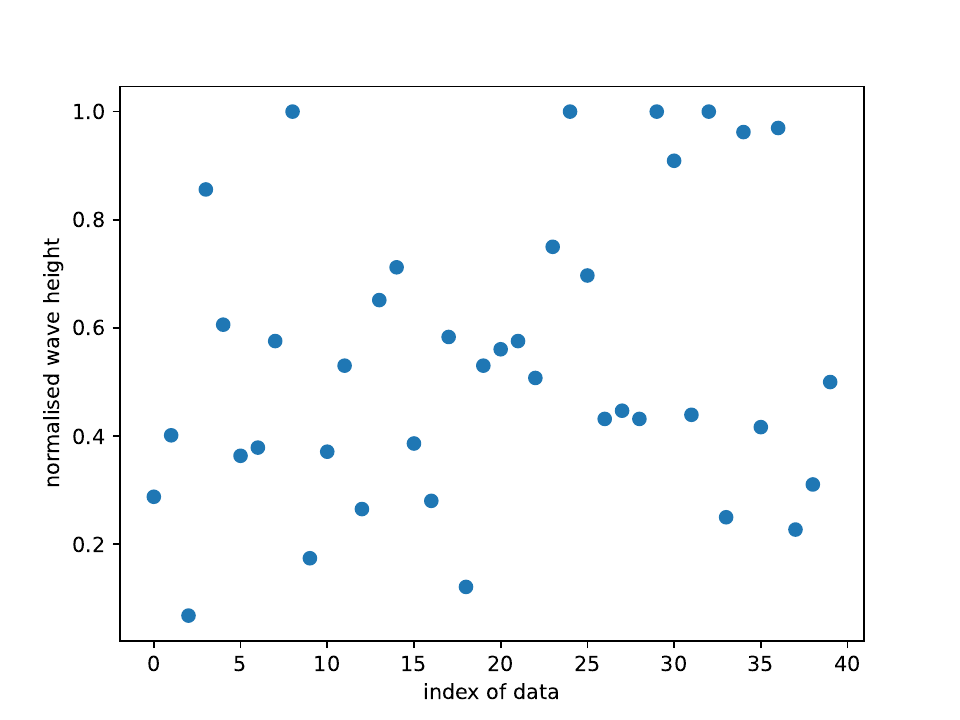}}}
\subfigure[Level 3 training data]{
\resizebox{0.3\textwidth}{!}{
\includegraphics[scale=0.8]{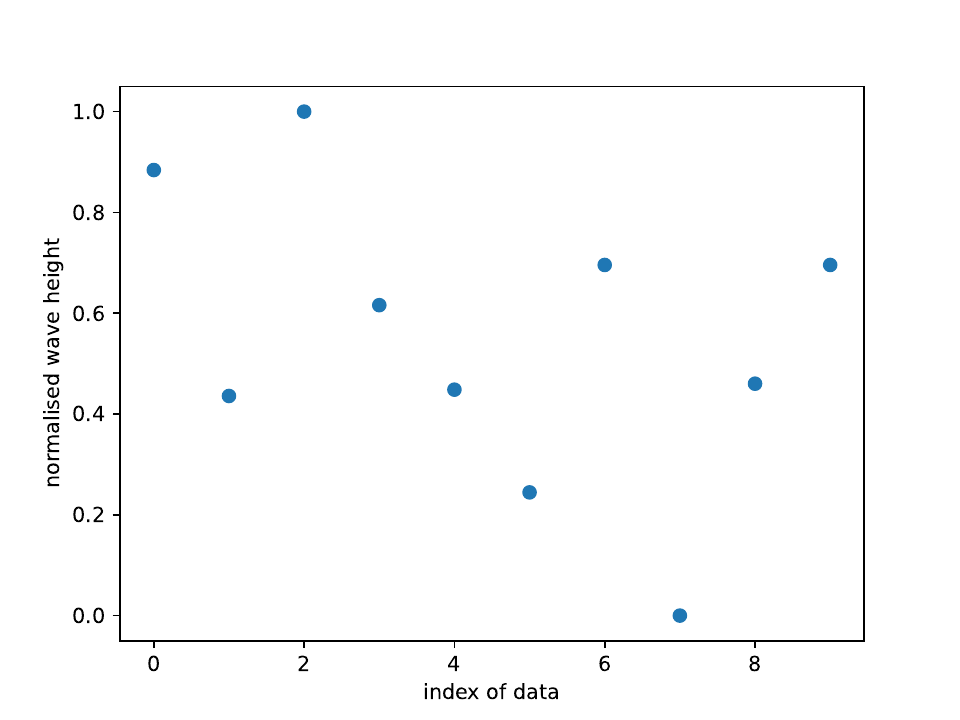}}}
\end{center}
\end{figure}

We generated the input design using Latin Hypercube Sampling (LHS), with $60$, $40$, and $10$ samples at levels 1, 2, and 3, respectively, where the sample sizes are chosen to represent the limited budgets typically available for computationally demanding simulators; the resulting outputs are shown in Figure~\ref{fig:tsunamiData}, illustrating the smooth variation across the input space due to the underlying shallow water dynamics.

\subsubsection{Results}
\begin{figure}[!htb]
\caption{Leave-one-out validation of the five emulation methods for the tsunami model. Red points indicate test observations falling outside the $95\%$ predictive interval.\label{fig:modelComparisonForTsunami}}
\includegraphics[width=\textwidth]{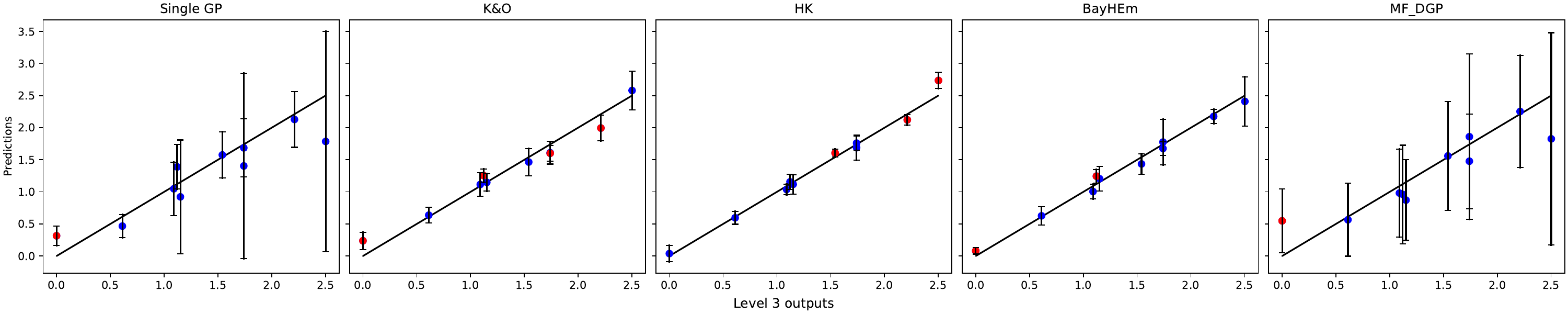}  
\end{figure}

The comparative performance of the five emulation methods on the tsunami model is presented in Table~\ref{tab:resOfTsunami} and Figure~\ref{fig:modelComparisonForTsunami}.
\begin{table}[h]
    \centering
    \caption{Comparative performance of different emulation methods on the tsunami model. Lower NRMSE indicates better accuracy, higher SCORE indicates better overall performance, and Coverage probability closer to the nominal value (0.95) indicates better uncertainty quantification. }
    \begin{tabular}{|c|c|c|c|c|c|}\hline
      \textbf{Metric} & \textbf{Single GP} & \textbf{K\&O} & \textbf{HK} & \textbf{BayHEm} & \textbf{MF\_DGP} \\ \hline
      NRMSE  & 0.119 & 0.052 & 0.035 & 0.031 & 0.124 \\ \hline
      Score & -0.984 & 1.853 & 2.693 & 3.025 & 1.111 \\ \hline
      Coverage probability & 90\% & 60\% & 70\% & 80\% & 90\% \\ \hline
    \end{tabular} 
    \vspace{0.5cm}    
    \label{tab:resOfTsunami}
\end{table}

\subsection{Application to the Joint UK Land Environment Simulator (JULES)}
\subsubsection{The JULES model}

JULES is a land surface model that represents hydrometeorological and biogeophysical processes, including the exchange of CO\textsubscript{2}, heat, and moisture between the land and atmosphere, as well as nutrient flows between vegetation and soils \citep{baker2022emulation}. We construct a two-fidelity JULES simulator focusing on a single grid cell representing the Netherlands, examining how broadleaf trees—one of the model's five plant functional types—influence the annual Gross Primary Productivity (GPP). GPP quantifies plant photosynthesis and marks the entry point of carbon into the terrestrial carbon cycle, making it a natural starting point for our analysis. The low-fidelity model uses 11 input parameters, while the high-fidelity incorporates two additional parameters governing the rate of leaf growth and the maximum leaf area; the high-fidelity runtime is twice that of the low-fidelity. The input parameters and their ranges are given in Table~\ref{tab:inputVar}, and the forcing data variables used by both fidelities are listed in Table~\ref{tab:forcingVar}.
 \begin{table}[h]
    \centering
     \caption{The input parameters for the broadleaf trees (BT) plant functional type, including their "standard" values and initial untuned ranges.}
    \begin{tabular}{lrrl}\toprule
      Parameter & Standard value & Range & Description \\ \midrule
      \text{alpha}  & $0.08$& $[0.04, 0.12]$ & \text{Quantum efficiency of photosynthesis (mol CO\textsubscript{2} (mol PAR photons)$^{-1}$)}\\
      \text{knl}  & $0.2$& $[0.05, 0.35]$ & \text{Rate of decay of N through the canopy}\\
      \text{g\_leaf\_0}  & $0.25$ & $[0.1, 3]$ & \text{Minimum turnover rate for leaves (360 d)$^{-1}$}\\
      \text{dqcrit(BT)} & $0.09$ & $[0.045, 0.18]$ & \text{Critical humidity deficit (kg H\textsubscript{2}O per kg air)}\\
      \text{f0(BT)} & $0.875$ & $[0.65, 0.972]$ & \text{Ci/Ca when dq = 0}\\
      \text{g\_grow(BT)}  & $20$& $[10, 40]$ & \text{Rate of leaf growth (360 d)$^{-1}$}\\
      \text{lai\_max(BT)}  & $ 7 $& $[3.5, 10]$ & \text{Maximum leaf area}\\
      \text{nmass(BT)}  & $0.0257$& $[0.0089, 0.0354]$ & \text{Top leaf N content (kgN per kgLeaf)}\\
      \text{rootd\_ft(BT)}  & $2$& $[0.1, 5.33]$ & \text{Parameter for decay of root functioning with depth (m)}\\
      \text{tleaf\_of(BT)}  & $ 278.15$& $[273, 283]$ & \text{Temperature below which leaves are dropped (K)}\\
      \text{tlow(BT)}  & $0$& $[-1, 1]$ & \text{Lower temperature parameter for photosynthesis ($^{\circ}\mathrm{C}$)}\\
      \text{tupp(BT)}  & $32$& $[22, 36]$ & \text{Upper temperature parameter for photosynthesis ($^{\circ}\mathrm{C}$)}\\
      \text{vsl(BT)}  & $32.50$& $[6, 150]$ & \text{Regression slope between $V_{\text{cmax}}$ and $N_{\text{area}}$ ($\mu$mol CO\textsubscript{2}gN$^{-1}$ s$^{-1}$)}\\
      \bottomrule\
    \end{tabular}  
    \label{tab:inputVar}
\end{table}

 \begin{table}[h]
    \centering
     \caption{The forcing data variables considered in the analysis.}
    \begin{tabular}{cc}\toprule
      Short Form Name & Long Form Name \\ \midrule
      \text{shortWav}  & \text{shortwave radiation} \\
      \text{longWav}  & \text{longwave radiation} \\
      \text{rainRate}  & \text{rain rate} \\
      \text{airTemp}  & \text{air temperature} \\
      \text{windSpeed}  & \text{wind speed} \\
      \text{pressure}  & \text{air pressure} \\
      \text{humidity}  & \text{specific humidity} \\
      \bottomrule\
    \end{tabular}  
    \label{tab:forcingVar}
\end{table}

\subsubsection{Data}
\begin{figure}[!thb]
\begin{center}
\caption{Two-level JULES model outputs.   \label{fig:julesData}}
\subfigure[Level 1 training data]{
\resizebox{0.45\textwidth}{!}{
\includegraphics[scale=0.8]{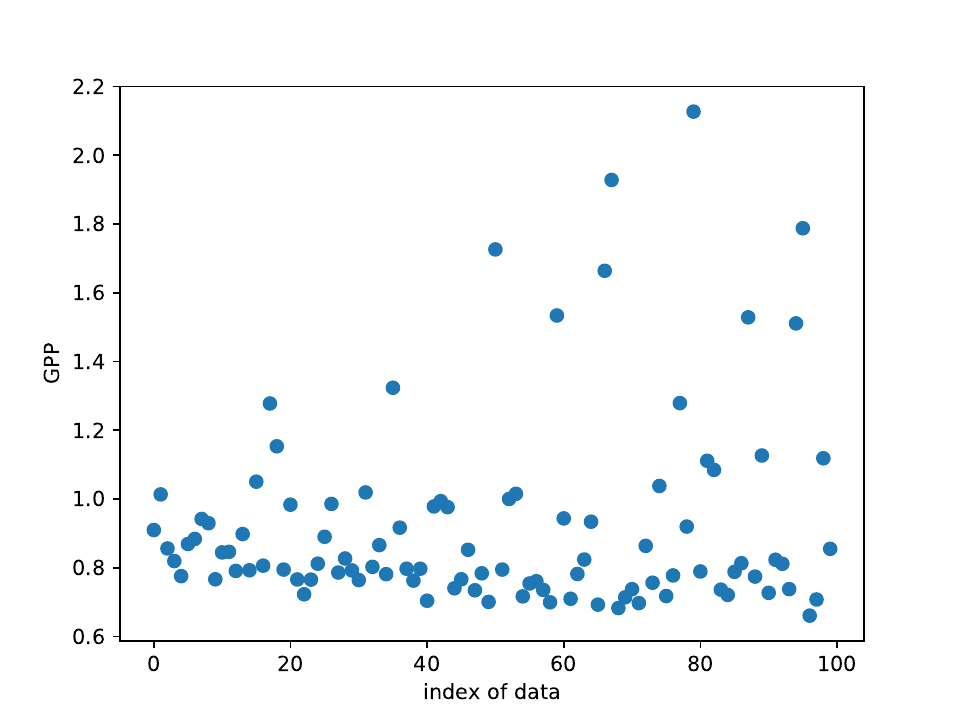}}} 
\subfigure[Level 2 training data]{
\resizebox{0.45\textwidth}{!}{
\includegraphics[scale=0.8]{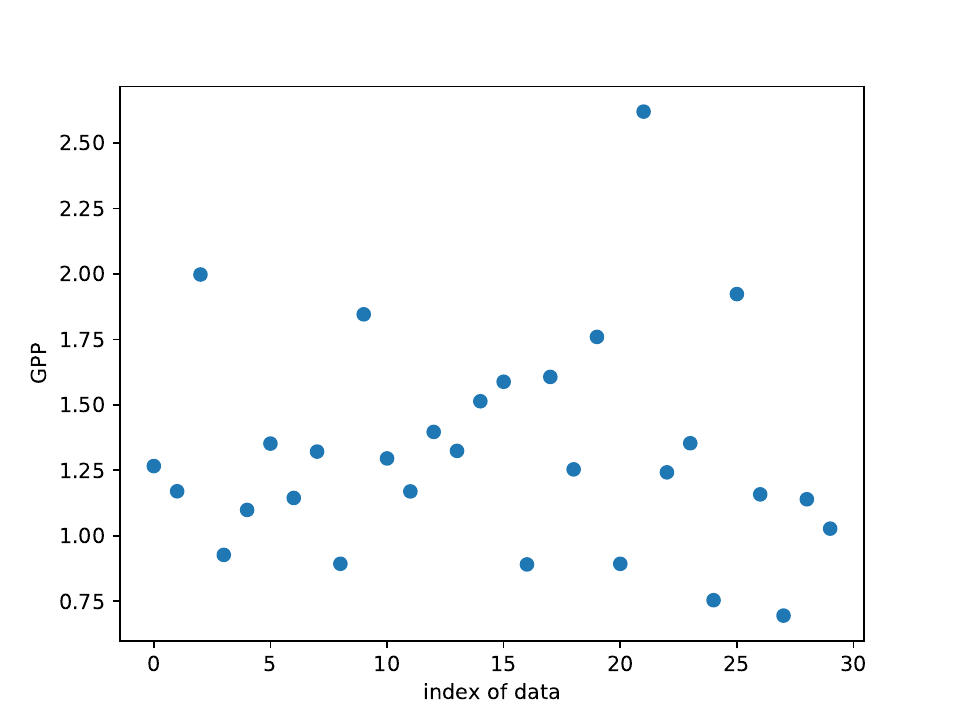}}}
\end{center}
\end{figure}

We generated the input design using Latin Hypercube Sampling across the input space defined in Table~\ref{tab:inputVar}, with $100$ and $30$ samples for the low- and high-fidelity models, respectively, adopting the same limited-budget rationale as in the tsunami application but with larger sample sizes to account for the higher input dimensionality ($13$ dimensions compared to $2$). Inspired by the nested design principle of \citet{qian2009nested}, we approximate a nested structure by matching each high-fidelity sample to its nearest low-fidelity neighbour in the shared 11-dimensional feature space. The matched low-fidelity inputs were then augmented with two additional high-fidelity features to form the complete 13-dimensional high-fidelity design. This matching strategy ensures that responses from both fidelity levels are available at approximately the same input locations, enabling direct estimation of the fidelity bias. The resulting outputs are presented in Figure~\ref{fig:julesData}, which illustrates the non-stationary behaviour across the parameter space.

\FloatBarrier
\subsubsection{Results}
\begin{figure}[!htb]
\caption{Leave-one-out validation of the five models for the JULES data. Red points indicate test observations falling outside the $95\%$ predictive interval. \label{fig:modelComparisonForJULES}}
\includegraphics[width=\textwidth]{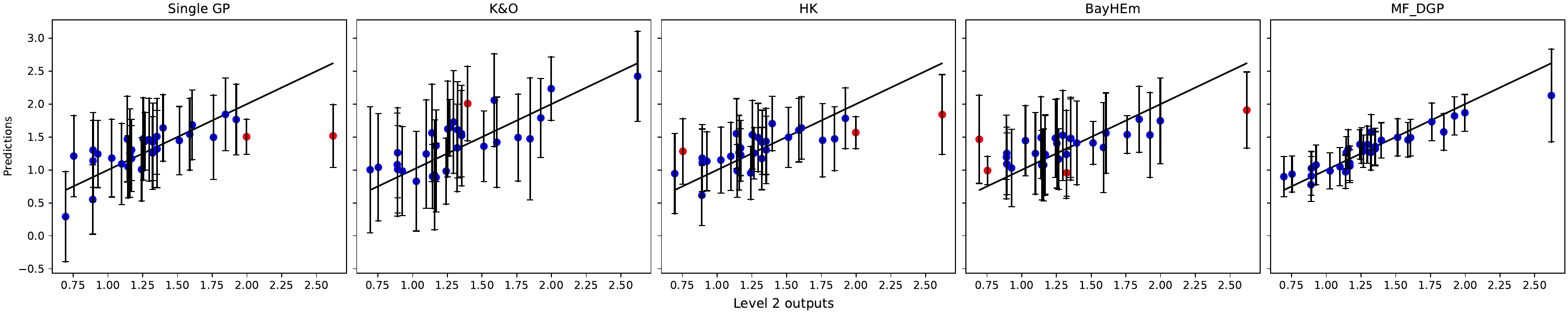}  
\end{figure}

The comparative performance of the five emulation methods on the JULES model is presented in Table~\ref{tab:resOfJules} and Figure~\ref{fig:modelComparisonForJULES}.
\begin{table}[h]
    \centering
    \caption{Comparative performance of different emulation methods on the JULES model. Lower NRMSE indicates better accuracy, higher SCORE indicates better overall performance, and Coverage probability closer to the nominal value (0.95) indicates better uncertainty quantification.}
    \begin{tabular}{|c|c|c|c|c|c|}\hline
      \textbf{Metric} & \textbf{Single GP} & \textbf{K\&O} & \textbf{HK} & \textbf{BayHEm} & \textbf{MF\_DGP} \\ \hline
      NRMSE  & 0.159 & 0.149 & 0.142 & 0.147 & 0.079 \\ \hline
      Score & 1.233 & 1.378 & 1.448 & 1.400 & 3.032 \\ \hline
      Coverage probability & 93\% & 97\% & 90\% & 87\% & 100\% \\ \hline
    \end{tabular} 
    \vspace{0.5cm}
    \label{tab:resOfJules}
\end{table}

\subsection{Results and Discussion}

Tables~\ref{tab:resOfTsunami} and~\ref{tab:resOfJules} summarise the leave-one-out performance of the five emulators, with the corresponding predictions shown in Figures~\ref{fig:modelComparisonForTsunami} and~\ref{fig:modelComparisonForJULES}. The two applications produce markedly different rankings, illustrating that the value of additional model flexibility depends on both the simulator and the amount of high-fidelity information available.

\paragraph{Tsunami application}
For the tsunami simulator, BayHEm gives the best performance according to both NRMSE (0.031) and SCORE (3.025). HK is close in mean predictive accuracy (NRMSE 0.035) but has lower empirical coverage (70\% compared with 80\% for BayHEm). K\&O also improves substantially on the single-fidelity GP in NRMSE, but its empirical coverage is only 60\%. MF-DGP, by contrast, has an NRMSE of 0.124, slightly worse than the single-fidelity GP (0.119), although both achieve 90\% empirical coverage.

These results are consistent with the tsunami application favouring comparatively parsimonious multi-fidelity structures. However, because the highest-fidelity design contains only 10 points, the empirical coverage values are necessarily coarse and should not be interpreted as precise estimates of long-run calibration. Likewise, the poor MF-DGP performance in this application is consistent with the difficulty of estimating a more flexible model from very limited high-fidelity information, but the present experiment alone does not identify a general sample-size threshold at which MF-DGP will succeed or fail.

\paragraph{JULES application}
For JULES, MF-DGP performs substantially better in mean prediction than the other approaches, with NRMSE 0.079 compared with 0.142--0.149 for the three traditional multi-fidelity methods and 0.159 for the single-fidelity GP. It also obtains the highest SCORE (3.032). All 30 held-out high-fidelity observations lie within the nominal 95\% predictive intervals, giving an empirical coverage of 100\% in this experiment. This should be interpreted as strong coverage on the available validation set rather than as evidence of perfect probabilistic calibration; intervals that are conservative can also produce 100\% empirical coverage in a finite sample.

The remaining multi-fidelity methods provide more modest improvements in NRMSE over the single-fidelity GP. K\&O has empirical coverage of 97\%, whereas HK and BayHEm obtain 90\% and 87\%, respectively. The strong MF-DGP performance is consistent with the possibility that the relationship between the two JULES fidelities is more difficult to represent using the simpler hierarchical constructions. The current experiment does not, however, isolate nonlinearity or non-stationarity as the unique cause of the performance difference; doing so would require dedicated diagnostics or controlled simulation experiments.

\paragraph{Comparison across applications}
Taken together, the case studies do not support a universally best multi-fidelity emulator. BayHEm performs particularly well for the tsunami example, whereas MF-DGP performs best for JULES. The contrast suggests that emulator complexity should be considered jointly with the structure of the inter-fidelity relationship and the amount of high-fidelity data. It also illustrates why predictive accuracy and uncertainty representation should be assessed separately: a method with a low NRMSE need not have empirical coverage close to the nominal level, and conversely high coverage alone does not guarantee accurate or well-calibrated predictions.

\section{Discussion}

This study compares four GP-based strategies for transferring information across fidelity levels in two Earth system applications. The main empirical finding is application dependence: relatively simple hierarchical methods perform strongly for the tsunami simulator, while the more flexible MF-DGP is clearly favoured for JULES according to NRMSE and SCORE. This finding argues against selecting a multi-fidelity method solely on the basis of model complexity or general methodological preference.

\subsection{Uncertainty propagation and predictive performance}
The methods differ importantly in how information from lower fidelities enters higher-fidelity predictions. HK uses the lower-fidelity posterior mean as a trend for the next level, whereas BayHEm propagates both posterior mean and covariance. K\&O uses an autoregressive relationship with a discrepancy process, and MF-DGP propagates uncertainty through a variational deep architecture. These structural differences provide plausible explanations for some of the observed accuracy--coverage trade-offs, but the two applications are not sufficient to establish a general causal ranking of the uncertainty-quantification properties of the methods.

In particular, empirical coverage should be interpreted in the context of sample size. For the tsunami application, one held-out point corresponds to 10 percentage points of coverage, while for JULES it corresponds to approximately 3.3 percentage points. Consequently, differences such as 70\% versus 80\% in the tsunami case should be treated as descriptive evidence rather than precise estimates of calibration. SCORE complements coverage by incorporating both predictive error and predictive variance, but neither metric alone establishes that a predictive distribution is calibrated in repeated use.

\subsection{Computational and design considerations}
Method selection also depends on simulator cost and experimental design. K\&O is most naturally implemented with nested or overlapping designs when discrepancy processes are fitted to paired fidelity evaluations. HK, BayHEm, and MF-DGP can be used with non-nested designs, which can be advantageous when simulations already exist at different locations or when designs are generated sequentially. The computational-complexity expressions in Table~\ref{tab:computational_complexity} should be viewed as leading-order comparisons: actual wall-clock cost depends on implementation, optimisation, the number of inducing points and Monte Carlo samples, and convergence criteria. For expensive simulators, emulator training cost may nevertheless be small relative to the cost of generating additional high-fidelity runs.

\subsection{Limitations and future directions}
Several limitations qualify the conclusions. First, only two applications are considered, with 10 and 30 high-fidelity simulations, respectively. The observed difference in MF-DGP performance is suggestive of an interaction between model flexibility and data availability, but it does not establish a universal minimum high-fidelity sample size. A systematic simulation study varying sample size while holding the underlying fidelity relationship fixed would be needed to quantify such a transition.

Second, the two applications differ simultaneously in input dimension, number of fidelity levels, simulator physics, design, and sample size. The present comparison therefore cannot attribute the change in method ranking uniquely to nonlinearity or non-stationarity. Future work could use controlled benchmark functions with known inter-fidelity relationships to separate these effects.

Third, the validation focuses on scalar outputs and leave-one-out predictions at the highest fidelity. Many Earth system applications require emulation of spatial fields, time series, or multiple correlated outputs. Extending the comparison to structured outputs, larger validation sets, and independent test designs would provide stronger evidence about generalisability and uncertainty calibration. Finally, empirical wall-clock timings and sensitivity to optimiser settings, inducing-point choices, and kernel specification would complement the theoretical computational comparison.

\FloatBarrier
\section{Conclusions}
The two case studies demonstrate that multi-fidelity emulation can substantially improve prediction of expensive high-fidelity Earth system simulators, but they also show that no single approach is uniformly superior. For the tsunami application, BayHEm achieves the lowest NRMSE and highest SCORE, while HK gives similarly accurate mean predictions. For JULES, MF-DGP achieves substantially lower NRMSE and a higher SCORE than the other methods. These contrasting rankings highlight the importance of matching emulator structure to the application rather than assuming that greater model complexity will necessarily improve performance.

The results also emphasise the need to evaluate predictive means and uncertainty estimates together. Empirical coverage varies substantially among methods, particularly in the small tsunami validation set, and high coverage should not by itself be interpreted as proof of calibration. The combination of NRMSE, SCORE, and coverage provides a more informative basis for comparison, while larger or independent validation designs would be needed for stronger calibration claims.

From a practical perspective, simpler hierarchical approaches remain attractive when high-fidelity simulations are very limited or when the relationship between fidelities can be represented adequately by a relatively simple structure. More flexible approaches such as MF-DGP may offer substantial gains when supported by sufficient data and when simpler structures fail to capture the relationship between fidelity levels. Future work should quantify these trade-offs systematically across sample sizes, fidelity structures, and higher-dimensional or structured Earth system outputs.

\section*{Data accessibility}
Code and data associated with this study are available at \url{https://github.com/XiaoyuHy/multifidelity-earth-system-emulation}.

\section*{Authors' contributions}
X.X.: conceptualisation, formal analysis, investigation, methodology, software, visualisation, writing--original draft, writing--review and editing; L.M.K.: formal analysis, investigation, methodology, software, visualisation, writing--review and editing; H.Y. and M.X.: data curation, investigation, software; J.M.S.: methodology, writing--review and editing; P.C.: methodology, writing--review and editing.

All authors gave final approval for publication and agreed to be held accountable for the work performed therein.

\section*{Conflict of interest declaration}
We declare we have no competing interests.

\section*{Funding}
This work was supported by the EPSRC ExCALIBUR programme through the Uncertainty Quantification at the Exascale (Exa-UQ) project, grant EP/W007886/1.

\bibliographystyle{copernicus}
\bibliography{multiLevelEmulation.bib}

\section*{Supplementary materials}
\subsection*{S1. Sparse Gaussian processes}
The standard Gaussian Process (GP) emulator, while effective for uncertainty quantification, encounters computational constraints when applied to large datasets due to its \(O(n^3)\) complexity for inference and \(O(n^2)\) storage requirements, where \(n\) denotes the number of data points. Additionally, for models with non-Gaussian likelihoods \(p(\bm{y} \mid \fv)\), where \(\bY\) represents observations and \(\fv\) denotes function values (as defined in Section \ref{sec:gpEmu}), direct analytical inference becomes intractable. To address these scalability and inference challenges, sparse Gaussian Processes (SGPs) with stochastic variational inference have been introduced, as in \citet{hensman2013gaussian}. A sparse GP approximates the full GP by introducing a set of \(M\) inducing points, denoted \(\induZ = (\bz_1, \dots, \bz_M)\), where \(M \ll n\), with corresponding latent function values \(\bu = f(\induZ)\). These inducing points act as a compact representation of the full GP, capturing essential information from the data while significantly reducing computational complexity. With the properties of GPs, the joint distribution of the full function values \(\fv\) and the inducing variables \(\bu\) is Gaussian:
\[
p(\fv, \bu) = p(\fv \mid \bu; \design, \induZ) p(\bu;\induZ),
\]
where \(p(\bu) = \mathcal{N}(\mean(\induZ), \Sigma(\induZ, \induZ^\prime))\) and \(p(\fv \mid \bu)\) is also Gaussian, with its mean and covariance obtained through standard Gaussian conditional identities. The joint density of the observations \(\bY\), the latent function values \(\fv\), and the inducing variables \(\bu\) can then be factorised as \( p(\bY, \fv, \bu) = p(\bY|\fv) p(\fv | \bu; \design, \induZ) p(\bu;\induZ)\).
To make inference computationally feasible, variational inference is used to approximate the true posterior \(p(\fv, \bu \mid \bY)\) by a simpler variational posterior \(q(\fv, \bu)\). The objective is to minimise the Kullback-Leibler (KL) divergence, \(\text{KL}[q \parallel p]\), between the variational posterior \(q\) and the true posterior \(p\), or equivalently, to maximise a lower bound \(\mathcal{L}\) on the log marginal likelihood of \(\bY\), given by:
\begin{equation} \label{eq:ELBO}
  \mathcal{L} = \mathbb{E}_{q(\fv,\bu)} \left[\log \frac{p(\bY,\fv,\bu)}{q(\fv, \bu)} \right].
\end{equation}
For computational efficiency, the variational posterior \(q(\fv, \bu)\) is typically chosen in the form:
\begin{align}
    &q(\fv, \bu) = p(\fv | \bu)q(\bu) \nonumber & \\
    &q(\bu) = \mathcal{N}(\bm{m}, \bS) \nonumber&
\end{align}
where \(q(\bu)\) is a Gaussian distribution parameterised by mean \(\bm{m}\) and covariance \(\bm{S}\), which are variational parameters optimised during inference. Using this choice of \(q(\fv, \bu)\), we can derive \(q(\fv)\) by marginalising out \(\bu\):
\begin{equation} \label{eq:posterior_f}
    q(\fv \mid \bm{m}, \bS;\design, \induZ) = \int p(\fv \mid \bu; \design, \induZ) q(\bu) \, d\bu.
\end{equation}
Since both \(p(\fv \mid \bu;\design, \induZ)\) and \(q(\bu)\) are Gaussian, \(q(\fv)\) is also Gaussian, with mean and covariance given by integrating over the variational distribution \(q(\bu)\). This choice of \(q(\fv, \bu)\) leads to a lower bound, often referred to as the evidence lower bound (ELBO), that can be expanded as:
\begin{equation}
  \mathcal{L} = \mathbb{E}_{q(\fv)}\left[ \log p(\bY|\fv) \right] - \text{KL}[q(\bu) \parallel p(\bu)],
\end{equation}
where the first term represents a data-fit term and the second term is the KL divergence between the approximate posterior \(q(\bu)\) and the prior \(p(\bu)\), which acts as a regulariser to keep \(q(\bu)\) close to the prior distribution. 
The sparse GP formulation reduces computational complexity for inference to \(O(M^2 n)\) and storage complexity to \(O(M n)\), where \(M \ll n\), making it feasible to scale GPs to larger datasets while retaining the essential properties of the original GP model. In summary, sparse GPs provide a scalable approach to GP modelling by leveraging inducing points and variational inference, enabling both efficient computation and flexibility in handling large datasets and non-Gaussian likelihoods.

\subsection*{S2. Deep Gaussian processes}

Deep Gaussian Processes (DGPs), as introduced by Damianou and Lawrence (2013), are hierarchical compositions of Gaussian processes that offer a powerful framework for flexible function approximation. Unlike single-layer GPs that are limited by kernel expressiveness, DGPs learn representation hierarchies non-parametrically with few hyperparameters. While single-layer GPs can be enhanced through learned or composite kernels, these approaches often risk overfitting or incur substantial computational costs. DGPs combine the flexibility of deep architectures with probabilistic predictions inherited from GP constructions, making them attractive for applications requiring both flexible function approximation and uncertainty estimation. 

A Deep Gaussian Process (DGP) defines a prior recursively on a sequence of vector-valued stochastic functions \( \deepF_1, \ldots, \deepF_L \). Each function \( \deepF_l \) is modelled as an independent GP across its dimensions. The structure is recursive: the outputs of one layer's GPs become the inputs to the next layer's GPs, with some Gaussian noise added between layers. More specifically, at each layer $l$, if $\deepF^l$ is \(D^l\)-dimensional, then it is modelled by \(D^l\) independent GPs, each taking the previous layer's outputs $\deepF^{l-1}$ as inputs. For a DGP with scalar outputs \(\bY\), the joint density is:
\begin{equation}\label{eq:jointDenDeepGP}
    p(\bY, \{\deepF^l, \deepU^l\}_{l=1}^L) = \prod_{i=1}^N p(y_i|f_i^L) \prod_{l=1}^L p(\deepF^l|\deepU^l; \deepF^{l-1}, \induZ^{l-1} )p(\deepU^l; \induZ^{l-1})
\end{equation}

where $\deepF^0 = \design$, $\deepU^l$ are inducing function values for each dimension, and $\induZ^{l-1}$ are inducing locations at each layer.

The key challenge in Deep Gaussian Processes (DGPs) has been achieving tractable inference. Earlier approaches, such as mean-field variational methods, imposed independence between layers, which significantly underestimated variance and resulted in poor performance in practice. Salimbeni and Deisenroth (2017) addressed this limitation using doubly stochastic variational inference. This approach preserves inter-layer correlations while maintaining computational efficiency by employing sparse approximations within each layer. Their method demonstrated effectiveness across datasets ranging in size from hundreds to billions of data points, consistently outperforming single-layer GPs without overfitting. In this framework, the posterior distribution over \(\{\deepU^l\}_{l=1}^L\) is assumed to factorise between both layers and dimensions. Consequently, the posterior takes the simple factorised form: 
\begin{equation}\label{eq:posteriorDeepGP}
    q(\{\deepF^l, \deepU^l\}_{l=1}^L) = \prod_{l=1}^L p(\deepF^l|\deepU^l; \deepF^{l-1}, \induZ^{l-1})q(\deepU^l)
\end{equation}

Similar to the single-layer SGP, the inducing variables at each layer can be marginalised analytically. Assuming \(q(\deepU^l)\) follows a Gaussian distribution with mean \(\bm{m}^l\) and covariance \(\bS^l\), this marginalisation yields a variational distribution that retains dependence through the layer-wise conditional structure:
\begin{equation}\label{eq:marginDeepU}
    q(\{\deepF^l\}_{l=1}^L) = \prod_{l=1}^L q(\deepF^l|\bm{m}^l, \bS^l; \deepF^{l-1}, \induZ^{l-1}) 
\end{equation} where \(q(\deepF^l|\bm{m}^l, \bS^l; \deepF^{l-1}, \induZ^{l-1})\) is defined as in (\ref{eq:posterior_f}). The evidence lower bound of the DGP is
\begin{equation} \label{eq:ELBOdgp}
  \mathcal{L}_{DGP} = \mathbb{E}_{q(\{\deepF^l, \deepU^l\}_{l=1}^L)} \left[\log \frac{p\left(\bY,\{\deepF^L,\deepU^L\}_{l=1}^L\right)}{ q(\{\deepF^l, \deepU^l\}_{l=1}^L)} \right].
\end{equation}
By substituting (\ref{eq:jointDenDeepGP}) and (\ref{eq:posteriorDeepGP}) into the corresponding expressions in (\ref{eq:ELBOdgp}) and performing some re-arrangement, we derive: 
\begin{equation}
  \mathcal{L}_{DGP} = \sum_{i=1}^N \mathbb{E}_{q(f_i^L)}[\log p\left(y_i|f_i^L\right)] - \sum_{l=1}^L \text{KL}[q(\deepU^l)||p\left(\deepU^l;\induZ^{l-1}\right)], 
\end{equation} where we leverage the exact marginalisation of the inducing variables (\ref{eq:marginDeepU}) and the property that each marginal distribution at the final layer depends only on the corresponding marginals from all preceding layers:
\begin{equation}
    q(f_i^L) = \int \prod_{l=1}^{L-1} q(f_i^l|\bm{m}^l, \bS^l; f_i^{l-1}, \induZ^{l-1})df_i^l
\end{equation}
To make predictions at a test location $\bx_*$, we first draw samples from the variational posterior by replacing the input locations with $\bx_*$. Denoting the function values at the test location as $f_*^l$, we can approximate the predictive density over $f_*^L$ using a Gaussian mixture:
\begin{equation}
    q(f_*^L) \approx \frac{1}{S}\sum_{s=1}^S q(f_*^L|\bm{m}^L, \bS^L; {f_{*}^{(s)}}^{L-1}, \induZ^{L-1})
\end{equation}
where $S$ denotes the number of samples of ${f_{*}^{(s)}}^{L-1}$.

\subsection*{S3. Multi-fidelity deep Gaussian processes}

MF-DGP extends the DGP framework of Section S2 to the multi-fidelity setting. The layer-wise sparse approximation introduces the variational posterior:
\begin{equation}
q(f^t_l|\mathbf{u}_l) = p(f^t_l|\mathbf{u}_l; \{f^t_{l-1}, \mathbf{X}_t\}, \mathbf{Z}_{l-1})q(\mathbf{u}_l)
\end{equation}
where $q(\mathbf{u}_l)$ follows a Gaussian distribution $\mathcal{N}(\mathbf{u}_l|\boldsymbol{\mu}_l, \boldsymbol{\Sigma}_l)$.

Training proceeds by jointly optimising all fidelity levels through stochastic variational inference, maximising the evidence lower bound:
\begin{equation}
\mathcal{L}_{\text{MF-DGP}} = \sum_{t=1}^T \sum_{i=1}^{N_t} \mathbb{E}_{q(f^{i,t}_t)}[\log p(y^{i,t}|f^{i,t}_t)] - \sum_{l=1}^L \text{KL}[q(\mathbf{u}_l) \| p(\mathbf{u}_l; \mathbf{Z}_{l-1})]
\end{equation}

The composite kernel at each intermediate layer takes the form:
\begin{align}
k_l &= k^{\rho}_l(\mathbf{x}_i, \mathbf{x}_j; \boldsymbol{\theta}^{\rho}_l)[{\sigma^2_l f^*_{l-1}(\mathbf{x}_i)^\top f^*_{l-1}(\mathbf{x}_j)} + k^{f-1}_l(f^*_{l-1}(\mathbf{x}_i), f^*_{l-1}(\mathbf{x}_j); \boldsymbol{\theta}^{f-1}_l)] \nonumber \\
&+ k^{\delta}_l(\mathbf{x}_i, \mathbf{x}_j; \boldsymbol{\theta}^{\delta}_l)
\end{align}
This structure explicitly incorporates both linear and nonlinear components of the inter-fidelity relationship and the correlation structure in the original input space.

Predictions at new locations $\mathbf{x}_*$ are obtained by propagating Monte Carlo samples recursively through the layers:
\begin{equation}
q(f^*_t) \approx \frac{1}{S}\sum_{s=1}^S q(f^{*}_t|\boldsymbol{\mu}_t, \boldsymbol{\Sigma}_t; \{f^{s,*}_{t-1}, \mathbf{x}_*\}, \mathbf{Z}_{t-1})
\end{equation}
The resulting computational complexity is $\mathcal{O}(SNM^2L)$, where $S$ is the number of Monte Carlo samples, $N$ the total number of observations, $M$ the number of inducing points, and $L$ the number of layers.

\end{document}